\documentclass[prl,preprint]{revtex4}
\usepackage{amssymb}
\usepackage{bm}
\usepackage{slashed}

\begin{document}

\title{Dark matter candidate with well-defined mass and couplings}
\author{Roland E. Allen and Aritra Saha}
\affiliation{Department of Physics and Astronomy \\
Texas A\&M University, College Station, Texas 77843, USA}

\begin{abstract}
We propose a Higgs-related but spin $1/2$ dark matter candidate with a mass
that is comparable to that of the Higgs. This particle is a
WIMP with an R-parity of $-1$, but it can be distinguished from a
neutralino by its unconventional couplings to W and Z bosons. 
Charged spin 1/2 particles of a new kind are also predicted at higher energy.
\end{abstract}

\maketitle

Although there is as yet no confirmed and statistically significant evidence
for direct, indirect, or collider-based detection of dark matter~\cite{drees}, 
all of these experiments are entering regimes where there is now a
reasonable possibility of success~\cite
{bertone,snowmass1,snowmass2,snowmass3,freese,lisanti,cirelli,klasen,undagoitia,fermi,ams,kahlhoefer,bergstrom}
. Weakly interacting massive particles (WIMPs) are among the leading dark
matter candidates, largely because they would have been created in about the
right abundance as thermal relics if their mass is $\sim 100$ GeV.

A favorite hypothetical WIMP is the lowest-mass neutralino, a linear
combination of the neutral fermionic superpartners predicted by
supersymmetry (susy). However, the tension that currently exists between
experiment and simple supersymmetric models may indicate that it is
desirable to consider alternative scenarios for how WIMPs can naturally
arise.

Here we propose a new candidate which resembles a neutralino, in that it has
spin 1/2 and is made stable by having an R-parity of -1, but it is
distinguished by various unusual features, including unconventional
couplings to the W and Z bosons and a well-defined mass that is simply
related to that of the recently discovered Higgs boson. One of the
motivations for this paper is in fact the Higgs discovery~\cite{ATLAS,CMS}.
In the present theory, the Higgs boson of fundamental physics is somewhat
analogous to the Higgs mode observed for superconductors in condensed matter
physics~\cite{varma}. Namely, the scalar Higgs field is interpreted as an
amplitude in a more complex structure described below.

We begin with an action that follows from fundamental arguments
given elsewhere~\cite{statistical}, but is here simply postulated as a
phenomenological model: 
\begin{eqnarray}
S_{\Phi }=S_{1}+S_{2}
\label{eq14.1}
\end{eqnarray}
where 
\begin{eqnarray}
S_{1}=\int d^{4}x\,\left( \frac{1}{2}\left( i\overline{\sigma }^{\mu
}D_{\mu }\Phi \left( x\right) \right) ^{\dag }\,\left( i\sigma ^{\nu }D_{\nu
}\Phi \left( x\right) \right) +h.c.\right) 
\label{eq14.2}
\end{eqnarray}
and $S_{2}$ consists of mass terms, discussed below, which result from
potentially very complicated field interactions in the complete potential
for Higgs scalars. Here $h.c.$ means Hermitian conjugate, the $\sigma $
matrices have their usual definitions (and are always implicitly multiplied
by an appropriate identity matrix), and $D_{\mu }$ is the usual covariant
derivative for the electroweak gauge fields.

Stripped of its fundamental derivation, this action still has a strong 
\textit{a posteriori} motivation: It leads to a dark matter candidate with
about the right abundance, which has a well-defined mass and well-defined
couplings, which is in a desirable mass range for direct detection, which is in the accessible energy range at the LHC, and whose
annihilation products are well-defined for indirect detection.

As described below, $\Phi $ consists of two Higgs-like doublets $\Phi
_{\uparrow }$ and $\Phi _{\downarrow }$, with each doublet having a neutral
and a charged component. This is reminiscent of the simplest description in general 
models with multiple electroweak Higgs doublets~\cite{2-doublets}. But here each of these four
components is itself a 2-component spinor, and scalar Higgs bosons are
interpreted as amplitudes of combinations of these spinors. This
very unusual picture is discussed below. The essential idea is that the
underlying structure involves spin 1/2 bosons, which are permitted because
there is a violation of Lorentz invariance, in this high-energy and
previously unexplored sector of the theory, which is fully consistent with
existing tests of Lorentz invariance~\cite{liberati,kostelecky}. Rotational invariance is still exact,
but invariance under a boost is not.

With  
\begin{eqnarray}
\Phi =\left( 
\begin{array}{c}
\Phi _{\uparrow } \\ 
\Phi _{\downarrow }
\end{array}
\right) 
\label{eq14.3}
\end{eqnarray}
it is convenient to use the same Weyl representation as for Dirac fields,
where 
\begin{eqnarray}
\gamma ^{\mu }=\left( 
\begin{array}{cc}
0 & \sigma ^{\mu } \\ 
\overline{\sigma }^{\mu } & 0
\end{array}
\right) \;,
\label{eq14.4}
\end{eqnarray}
so that (after integration by parts with neglect of boundary terms) 
\begin{eqnarray}
S_{1} &=&\int d^{4}x\,\frac{1}{2}\left( \Phi _{\uparrow }^{\dag }\left(
x\right) i\sigma ^{\mu }D_{\mu }\,i\overline{\sigma }^{\nu }D_{\nu }\Phi
_{\uparrow }\left( x\right) +\Phi _{\downarrow }^{\dag }\left( x\right) i
\overline{\sigma }^{\mu }D_{\mu }\,i\sigma ^{\nu }D_{\nu }\Phi _{\downarrow
}\left( x\right) \right) +h.c.  \label{eq7.52x} \\
&=&\int d^{4}x\,\left( -\frac{1}{2}\Phi ^{\dag }\left( x\right) \gamma ^{\mu
}D_{\mu }\,\gamma ^{\nu }D_{\nu }\Phi \left( x\right) \right) +h.c. \\
&=&-\int d^{4}x\,\frac{1}{2}\Phi ^{\dag }\left( x\right) \,\slashed{D}
^{2}\,\Phi \left( x\right) +h.c.
\label{eq14.5}
\end{eqnarray}

According to a result~\cite{schwartz} that can easily be extended to the
nonabelian case, we have 
\begin{eqnarray}
\slashed{D}^{2}=-D^{\mu }D_{\mu }+S^{\mu \nu }F_{\mu \nu }
\label{eq14.6}
\end{eqnarray}
with a $(-+++)$ convention for the metric tensor. The second term
gives an addition to standard physics, involving the field strength tensor $
F_{\mu \nu }$ for the electroweak gauge fields and the Lorentz generators $S^{\mu
\nu }$ which act on Dirac spinors:
\begin{eqnarray}
\mathcal{S}_{1}=\int d^{4}x\,\left( \frac{1}{2}\Phi ^{\dag }\left( x\right)
D^{\mu }D_{\mu }\Phi \left( x\right) -\frac{1}{2}\Phi ^{\dag }\left(
x\right) \,S^{\mu \nu }F_{\mu \nu }\,\Phi \left( x\right) \right) +h.c.
\label{eq14.7}
\end{eqnarray}
where 
\begin{eqnarray}
S^{\mu \nu }=\frac{1}{2}\sigma ^{\mu \nu }\;
\end{eqnarray}
or \cite{schwartz} 
\begin{eqnarray}
S^{kk^{\prime }}=\frac{1}{2}\varepsilon _{kk^{\prime }k^{\prime \prime
}}\left( 
\begin{array}{cc}
\sigma ^{k^{\prime \prime }} & 0 \\ 
0 & \sigma ^{k^{\prime \prime }}
\end{array}
\right) \quad ,\quad S^{0k}=-\frac{i}{2}\left( 
\begin{array}{cc}
\sigma ^{k} & 0 \\ 
0 & -\sigma ^{k}
\end{array}
\right) \;\;.
\label{eq14.8}
\end{eqnarray}

This can be rewritten in terms of the 
``magnetic'' and 
``electric'' fields $B_{k}$ and $E_{k}$ defined by 
\begin{eqnarray}
F_{kk^{\prime }}=-\varepsilon _{kk^{\prime }k^{\prime \prime }}B_{k^{\prime
\prime }}\quad ,\quad F_{0k}=E_{k}
\label{eq14.9}
\end{eqnarray}
since \cite{schwartz} 
\begin{eqnarray}
-S^{\mu \nu }F_{\mu \nu }=\left( 
\begin{array}{cc}
\left( \overrightarrow{B}+i\overrightarrow{E}\right) \cdot \overrightarrow{
\sigma } & 0 \\ 
0 & \left( \overrightarrow{B}-i\overrightarrow{E}\right) \cdot 
\overrightarrow{\sigma }
\end{array}
\right) 
\label{eq14.10}
\end{eqnarray}
where $a\cdot b=a_{k}b^{k}$ (with $\mu =0,1,2,3$ and $k=1,2,3$). We then
obtain 
\begin{eqnarray}
\mathcal{S}_{1}=\int d^{4}x\,\left( \Phi ^{\dag }\left( x\right) D^{\mu
}D_{\mu }\Phi \left( x\right) +\Phi ^{\dag }\left( x\right) \,
\overrightarrow{B}\cdot \overrightarrow{\sigma }\,\Phi \left( x\right)
\right) \;.
\label{eq14.11}
\end{eqnarray}
The second term (which is analogous to the interaction of an electron spin
with a magnetic field) is invariant under a rotation, but not under a boost,
making it the only aspect of the theory that does not have complete Lorentz
invariance. As discussed below, this term will have observable effects only
at high energy (or in extremely weak radiative corrections), and in conjunction
with the new spin 1/2 particles predicted here.

Let us regroup the components according to charge: 
\begin{eqnarray}
\Phi ^{r}=\left( 
\begin{array}{c}
\Phi _{\uparrow }^{r} \\ 
\Phi _{\downarrow }^{r}
\end{array}
\right) \;,\;r=0\text{ or }+
\label{eq14.12}
\end{eqnarray}
and then write
\begin{eqnarray}
\Phi ^{r}=\phi ^{r}\chi ^{r}\;\text{[no sum on }r\text{]}
\label{eq14.13}
\end{eqnarray}
where $\phi ^{r}$\ is a complex scalar and $\chi ^{r}$ is a 4-component
spinor. We can achieve a scalar condensate and scalar excitations by
requiring that
\begin{eqnarray}
\chi _{\downarrow }^{r\,\dag }\overrightarrow{\sigma }\,\chi _{\downarrow
}^{r}=-\chi _{\uparrow }^{r\,\dag }\overrightarrow{\sigma }\,\chi _{\uparrow
}^{r}\;\text{[no sum on }r\text{]}
\label{eq14.13a}
\end{eqnarray}
or
\begin{eqnarray}
\chi ^{r\,\dag }\overrightarrow{\sigma }\,\chi ^{r}=0\;.
\label{eq14.15}
\end{eqnarray}
As a bonus the unconventional term in (\ref{eq14.11}) also vanishes:   
\begin{eqnarray}
\Phi ^{\dag }\left( x\right) \,\overrightarrow{B}\cdot \overrightarrow{
\sigma }\,\Phi \left( x\right) =0\;.
\label{eq14.16}
\end{eqnarray}
With the normalization
\begin{eqnarray}
\chi ^{r\,\dag }\chi ^{r}=1\;\text{[no sum on }r\text{]}\;,
\label{eq14.17}
\end{eqnarray}
(\ref{eq14.11}) is then reduced to  
\begin{eqnarray}
\mathcal{S}_{1}=\int d^{4}x\,\phi ^{\dag }\left( x\right) D^{\mu }D_{\mu
}\phi \left( x\right) 
\label{eq14.14}
\end{eqnarray}
where $\phi $ has the scalar components $\phi ^{r}$. 
Each amplitude mode $\phi ^{r}$ thus has only its standard coupling to the
gauge fields through the covariant derivative, and we have returned to
standard physics with scalar Higgs bosons.

In the present theory, however, there can also be spin $1/2$ excitations, analogous to
quasiparticle excitations in a superconductor. They will be harder to create
than scalar Higgs bosons, because angular momentum conservation requires
that they be created in pairs. They will also be harder to observe, because
there are no apparent decay modes for the lowest-mass of these particles:
They have an R-parity of $-1$, with spin $1/2$ and with no lepton or baryon
number.

The masses of these particles depend on the potentially very complicated set
of parameters determining $S_{2}$, made even more nontrivial by the
requirements of susy~\cite{baer-tata,kane-susy}, which doubles the number of
Higgs fields yet again. But if the action is extremalized, by setting the
first derivatives of the full Higgs potential $V$\ equal to zero, and the scalar
mass eigenstates $\phi _{i}$ are then determined, by diagonalizing the mass
matrix obtained from the second derivatives of $V$, the resulting
lowest-order Lagrangian has the form
\begin{eqnarray}
-\mathcal{L}_{2}=\sum\limits_{i}m_{i}^{2}\phi _{i}^{\ast }\phi _{i}
\label{eq14.19}
\end{eqnarray}
which is equivalent to
\begin{eqnarray}
-\mathcal{L}_{2}=\sum\limits_{i}m_{i}^{2}\Phi _{i}^{\dag }\Phi _{i}
\label{eq14.20}
\end{eqnarray}
if we can still write $\Phi_{i}=\phi_{i}\chi _{i}$ for the mass eigenstates, with $\chi _{i}$ constant. More generally, with
\begin{eqnarray}
\Phi_{i} =\left( 
\begin{array}{c}
H _{i,\uparrow } \\ 
H _{i,\downarrow }
\end{array}
\right)  \; ,
\label{eq14.21}
\end{eqnarray}
we then have
\begin{eqnarray}
-\mathcal{L}_{2}=\sum\limits_{i}m_{i}^{2}\left( H _{i,\uparrow }^{\dag } H _{i,\uparrow } + H _{i,\downarrow }^{\dag } H _{i,\downarrow } \right) \; .
\label{eq14.22}
\end{eqnarray}
The masses for the
2-component spinors $H _{i}$ are then the same as the masses for the Higgs
scalars $\phi _{i}$ in this simplest description, and the couplings are also the same. 

In the simplest version of susy, with two Higgs
doublets $\phi _{u}$ and $\phi _{d}$ -- which correspond to $\Phi _{u}$ and $
\Phi _{d}$ in the present picture -- there are 3 would-be Goldstone bosons,
one charged Higgs, and 3 neutral Higgses, which then account for the 4
charged and 4 neutral degrees of freedom. The same basic approach is 
needed here, but there must be 3 charged and 7 neutral scalar 
Higgs bosons (resulting from the various scalar combinations of fields in $\Phi _{u}$ and $
\Phi _{d}$), and the same number of types of 
spin 1/2 particles, to account for the 8 charged and 8 neutral degrees of
freedom. Their masses must arise from the field 
interactions, including the variety of quartic terms that couple the fields to one 
another. 

According to the spin-statistics theorem, spin $1/2$ bosonic excitations are
impossible, but the requirements of this theorem are not satisfied in this
one specific context, since the second term of (\ref{eq14.11}) is not fully Lorentz invariant:
It is invariant under a rotation, but not a Lorentz boost with respect to
the original (cosmological) coordinate system.

The present theory is, however, fully Lorentz invariant if the internal
(spinor) degrees of freedom in $\Phi $ are not excited -- as can be seen in 
(\ref{eq14.14}) and (\ref{eq14.19}) -- and these excitations can be observed only at the high
energies that are now becoming available. Furthermore, the extremely weak
virtual effects of these excitations are irrelevant to the many existing
sensitive tests of Lorentz invariance, which probe only rotational
invariance (rather than boosts) or else those phenomena in various areas of
physics and astrophysics where the present theory is fully Lorentz 
invariant~\cite{liberati,kostelecky}. In particular, it is hard to imagine how 
any of the many experiments listed in Ref.~\cite{kostelecky} could have 
any measurable sensitivity to virtual processes involving pairs of 125 GeV spin 1/2 particles 
which only participate in the weak interaction, 
exactly obey rotational invariance, and have free-particle propagators 
(resulting from (\ref{eq14.11})  and  (\ref{eq14.20})) that are fully Lorentz invariant.

Two historical precedents may be relevant: After the electron was discovered
in 1897, and the photon was introduced by Einstein in 1905, the richness of
behavior associated with spin 1/2 fermions and spin 1 gauge bosons emerged
slowly during the following decades. (J.J. Thomson in 1897 did not picture
the electron as described by a 4-component Dirac field.) More than a century
later, the third kind of Standard Model particle, with spin 0, has finally
been discovered, and one should not be completely surprised if some of its
implications are yet to be determined. Similarly, it should not be
completely surprising if Lorentz invariance, like previously
well-established principles such as P and CP invariance, is ultimately found
to have exceptions in a more nearly complete theory.

We obtain no direct interaction of these new particles with fermions in the
present formulation, but they can be produced by
quarks in colliding protons, via virtual W or Z bosons.

Details of the phenomenology for the various kinds of experiments are of great interest, but beyond
the scope of the present paper. We only note that once a lowest-mass
particle of this kind has left the region where it was created, it is unable
to decay without violating lepton number or
baryon number conservation, since the net decay products must have angular
momentum $1/2$. This implies that these (weakly-interacting) particles are
dark matter candidates, roughly similar to neutralinos, but distinguished by
both their quite different couplings and the fact that their mass is simply
related to that of the recently discovered Higgs boson.

\end{document}